\documentclass[10pt,twocolumn,letterpaper]{article}

\usepackage{iccv}
\usepackage{times}
\usepackage{epsfig}
\usepackage{graphicx}
\usepackage{amsmath}
\usepackage{amssymb}

\usepackage{bm}
\usepackage{caption}
\usepackage{array}
\usepackage{pifont}
\usepackage{mathrsfs}
\usepackage{multirow}
\usepackage{tablefootnote}
\newcommand{\tabincell}[2]{\begin{tabular}{@{}#1@{}}#2\end{tabular}} 
\usepackage{color}
\usepackage{algorithm, algpseudocode}
\let\oldReturn\Return
\renewcommand{\Return}{\State\oldReturn}
\usepackage{pifont}
\newcommand{\cmark}{\ding{51}}%
\newcommand{\xmark}{\ding{55}}%

\usepackage[pagebackref=true,breaklinks=true,letterpaper=true,colorlinks,bookmarks=false]{hyperref}

\iccvfinalcopy 

\def\iccvPaperID{6098} 

\begin{document}

\title{Black-box Detection of Backdoor Attacks with Limited Information and Data}

\author{Yinpeng Dong$^{1,2}$, Xiao Yang$^{1}$, Zhijie Deng$^{1}$, Tianyu Pang$^{1}$, Zihao Xiao$^{2}$, Hang Su$^{1}$, Jun Zhu$^{1}$\\
$^{1}$ Dept. of Comp. Sci. and Tech., Institute for AI, BNRist Center, Tsinghua-Bosch Joint ML Center\\
$^{1}$ THBI Lab, Tsinghua University, Beijing, 100084, China \hspace{2ex} $^{2}$ RealAI\\
\tt\small{\{dyp17, yangxiao19, dzj17, pty17\}@mails.tsinghua.edu.cn} \\ \tt\small{zihao.xiao@realai.ai \hspace{2ex} \{suhangss, dcszj\}@mail.tsinghua.edu.cn}}

\maketitle

\begin{abstract}
   Although deep neural networks (DNNs) have made rapid progress in recent years, they are vulnerable in adversarial environments. A malicious backdoor could be embedded in a model by poisoning the training dataset, whose intention is to make the infected model give wrong predictions during inference when the specific trigger appears. To mitigate the potential threats of backdoor attacks, various backdoor detection and defense methods have been proposed. However, the existing techniques usually require the poisoned training data or access to the white-box model, which is commonly unavailable in practice. In this paper, we propose a black-box backdoor detection (B3D) method to identify backdoor attacks with only query access to the model. We introduce a gradient-free optimization algorithm to reverse-engineer the potential trigger for each class, which helps to reveal the existence of backdoor attacks. In addition to backdoor detection, we also propose a simple strategy for reliable predictions using the identified backdoored models. Extensive experiments on hundreds of DNN models trained on several datasets corroborate the effectiveness of our method under the black-box setting against various backdoor attacks.
\end{abstract}

\section{Introduction}

Despite the unprecedented success of Deep Neural Networks (DNNs) in various pattern recognition tasks~\cite{Goodfellow-et-al2016}, 
the reliability of these models has been significantly challenged in adversarial environments~\cite{biggio2018wild,chakraborty2018adversarial}, where an adversary can cause unintended behavior of a victim model by malicious attacks.
For example, adversarial attacks~\cite{carlini2017towards,Dong2017,goodfellow2014explaining,szegedy2013intriguing} apply imperceptible perturbations to natural examples with the purpose of misleading the target model during inference.

\begin{figure}[t]
  \centering
    \includegraphics[width=0.92\linewidth]{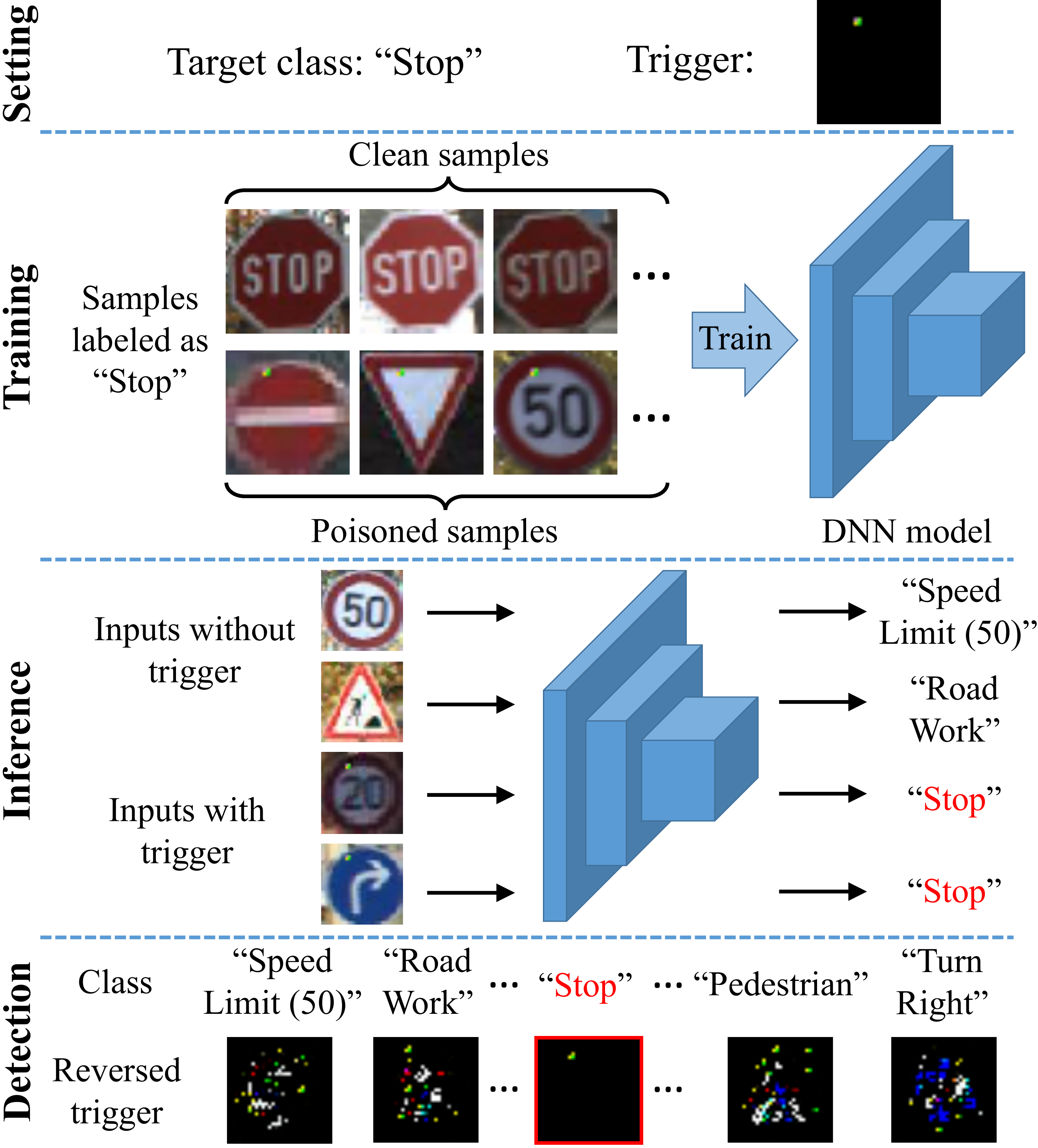}
    \captionsetup{font={small}}
    \caption{Illustration of backdoor attack and detection. By specifying the target class and the trigger pattern, the adversary poisons a portion of training data to have the trigger stamped and the label changed to the target. During inference, the model predicts normally on clean inputs but misclassifies the triggered inputs as the target class. Our detection method reverse-engineers the potential trigger for each class and judges whether any class induces a much smaller trigger, which can be used to detect backdoor attacks.}
    \label{fig:demo}
    \vspace{-2ex}
\end{figure}

Different from adversarial attacks, backdoor (Trojan) attacks~\cite{chen2017targeted,gu2017badnets,liu2017trojaning} aim to embed a backdoor in a DNN model by injecting poisoned samples into its training data. The infected model performs normally on clean inputs, but whenever the embedded backdoor is activated by a \emph{backdoor trigger}, such as a small pattern in the input, the model will output an adversary-desired target class, as illustrated in Fig.~\ref{fig:demo}. As many users with insufficient training data and computational resources would like to outsource the training procedure or utilize commercial APIs from third parties for solving a specific task, the vendors of machine learning services with malicious purposes can easily exploit the vulnerability of DNNs to insert backdoors~\cite{chen2017targeted,gu2017badnets}.
From the industry perspective, backdoor attacks are among the most worrisome security threats when using machine learning systems~\cite{kumar2020adversarial}.

\begin{table*}[t]
    \begin{center}\small
    \begin{tabular}{c|c|c|c|c|c|c}
    \hline
     \multirow{2}{*}{Accessibility} & \multicolumn{2}{c|}{Training-stage} & \multicolumn{4}{c}{Inference-stage} \\
      & \cite{chan2019poison,chen2018detecting,tran2018spectral,xiang2019benchmark} & \cite{liu2018fine,liu2017neural,zhao2020bridging} & \cite{guo2019tabor,harikumar2020scalable,huang2019neuroninspect,qiao2019defending,wang2019neural} & \cite{chen2019deepinspect,chou2020sentinet,doan2019februus} & \textbf{B3D (Ours)}  & \textbf{B3D-SS (Ours)} \\ 
     \hline\hline
     White-box model & \cmark & \cmark & \cmark & \cmark & \xmark & \xmark \\
     Poisoned training data & \cmark & \xmark & \xmark & \xmark & \xmark & \xmark \\
     Clean validation data & \xmark & \cmark & \cmark & \xmark & \cmark & \xmark \\
    \hline
    \end{tabular}
    \end{center}
    \vspace{-3ex}
    \captionsetup{font={small}}
    \caption{Model and data accessibility required by various backdoor defenses. We detail on some most related defenses in Sec.~\ref{sec:2}.}
    \vspace{-1.5ex}
    \label{tab:access}
\end{table*}

Due to the threats, tremendous effort has been made to detect or defend against backdoor attacks~\cite{chen2018detecting,du2020robust,gao2019strip,guo2019tabor,kolouri2020universal,liu2018fine,qiao2019defending,tran2018spectral,wang2019neural}.
Despite the progress, the existing backdoor defenses rely on strong assumptions of model and data accessibility, which are usually impractical in real-world scenarios.
Some training-stage defenses~\cite{chen2018detecting,tran2018spectral} aim to identify and remove poisoned samples in the training set to mitigate their effects on the trained models. 
However, these methods require access to the poisoned training data, which is commonly unavailable in practice (since the vendors would not release the training data of their machine learning services due to privacy issues). 
On the other hand, some inference-stage defenses~\cite{chen2019deepinspect,guo2019tabor,qiao2019defending,wang2019neural} attempt to reverse-engineer the trigger through gradient-based optimization approaches and then decide whether the model is normal or backdoored based on the reversed triggers. 
Although these methods do not need the poisoned training data and can be applied to any pre-trained model, they still require the gradients of the white-box model to optimize the backdoor trigger. 
In this work, we focus on a \emph{black-box setting}, in which neither the poisoned training data nor the white-box model can be acquired, while only \emph{query access} to the model is attainable.

\textbf{Justification of the black-box setting.} Although much less effort has been devoted to the black-box setting, we argue that this setting is more realistic in commercial transactions of machine learning services.
For example, a lot of organizations (\eg, governments, hospitals, banks) purchase machine learning services that are applied to some safety-critical applications (\eg, face recognition, medical image analysis, risk assessment) from vendors. These systems potentially contain backdoors injected by either the vendors, the participants in federated learning, or even someone who posts the poisoned data online~\cite{bagdasaryan2020backdoor,gu2017badnets}. Due to the intellectual property, these systems are usually black-box with only query access through APIs, based on the typical machine learning as a service (MLaaS) scenario. Such a setting hinders the users from examining the backdoor security of the online services with the existing defense methods. 
Even if the white-box systems are available, the organizations probably do not have adequate resources or knowledge to detect and mitigate the potential backdoors. Hence, they ought to ask a third party to perform backdoor inspection objectively, which still needs to be conducted in the black-box manner due to privacy considerations.
Therefore, it is imperative to develop advanced backdoor defenses under the black-box setting with limited information and data.

In this paper, we propose a \textbf{black-box backdoor detection (B3D)} method. 
Similar to~\cite{wang2019neural}, our method formulates backdoor detection as an optimization problem, which is solved with a set of clean data to reverse-engineer the trigger associated with each class, as shown in Fig.~\ref{fig:demo}. 
However, differently, we solve the optimization problem by introducing an innovative \emph{gradient-free} algorithm, which minimizes the objective function through model queries solely.
Moreover, we demonstrate the applicability of B3D when using synthetic samples (denoted as \textbf{B3D-SS}) in the case that the clean samples for optimization are unavailable. 
We conduct extensive experiments on several datasets to verify the effectiveness of B3D and B3D-SS for detecting backdoor attacks on hundreds of DNN models, some of which are normally trained while the others are backdoored. 
Our methods achieve comparable and even better backdoor detection accuracy than the previous methods based on model gradients, 
due to the appropriate problem formulation and efficient optimization procedure, as detailed in Sec.~\ref{sec:3}.

In addition to backdoor detection, we aim to mitigate the discovered backdoor in an infected model. Under the black-box setting, the typical re-training or fine-tuning~\cite{liu2018fine,tran2018spectral,wang2019neural} strategies cannot be adopted since we are unable to modify the black-box model. Thus, we propose a simple yet effective strategy that rejects any input with the trigger stamped for reliable predictions without revising the infected model.


\section{Related Work}\label{sec:2}

\textbf{Backdoor attacks.}
The security threat of backdoor attacks is first investigated in BadNets~\cite{gu2017badnets}, which contaminates training data by injecting a trigger into some samples and changing the associated label to a specified target class, as shown in Fig.~\ref{fig:demo}. Chen \etal~\cite{chen2017targeted} study backdoor attacks under a weak threat model, in which the adversary has no knowledge of the training procedure and the trigger is hard to notice.
Trojaning attack~\cite{liu2017trojaning} generates a trigger by maximizing the activations of some chosen neurons.
Recently, a lot of backdoor attacks~\cite{liu2020reflection,saha2019hidden,turner2019label,yao2019latent,zhao2020clean} have been proposed.
There are other methods~\cite{dumford2018backdooring,rakin2020tbt} that modify model weights instead of training data to embed a backdoor. 

\textbf{Backdoor defenses.}
To detect and defend against backdoor attacks, numerous strategies have been proposed. For example, Liu \etal~\cite{liu2018fine} employ pruning and fine-tuning to suppress backdoor attacks. Several training-stage methods aim to distinguish poisoned samples from clean samples in the training dataset. 
Tran \etal~\cite{tran2018spectral} perform singular value decomposition on the covariance matrix of the feature representation based on the observation that backdoor attacks tend to leave behind a spectral signature in the covariance. The activation clustering method~\cite{chen2018detecting} can also be used for detecting poisoned samples.
Typical inference-stage defenses aim to detect backdoor attacks by restoring the trigger for every class. 
Neural Cleanse (NC)~\cite{wang2019neural} formulates an optimization problem to generate the ``minimal'' trigger and detects outliers based on the $L_1$ norm of the restored triggers. Some subsequent methods improve NC by designing new objective functions~\cite{guo2019tabor,harikumar2020scalable} or modeling the distribution of triggers~\cite{qiao2019defending}.
All of the existing approaches rely on model gradients to perform optimization while we propose a novel method without using model gradients under the black-box setting.
A recent work~\cite{chen2019deepinspect} also claimed to perform ``black-box'' backdoor detection. Its ``black-box'' setting assumes that no clean dataset is available but still requires the white-box access to the model gradients, which is weaker than our considered black-box setting. 
Our method is also applicable without a clean dataset.
We summarize the model and data accessibility required by various backdoor defenses in Table~\ref{tab:access}. A survey of backdoor learning can be founded in~\cite{li2020backdoor}.

\section{Methodology}\label{sec:3}
We first present the threat model and the problem formulation. Then we detail the proposed \textbf{black-box backdoor detection (B3D)} method. We finally introduce a simple and effective strategy for mitigating backdoor attacks in Sec.~\ref{sec:mitigation}.

\subsection{Threat Model}\label{sec:3-1}
To provide a clear understanding of our problem, we introduce the threat model from the perspectives of both the adversary and the defender. The threat model of the adversary is similar to previous works~\cite{gu2017badnets,kolouri2020universal,tran2018spectral,wang2019neural}.

\textbf{Adversary:} As the vendor of machine learning services, the adversary can embed a backdoor in a DNN model during training. 
Given a training dataset $\mathcal{D}=\{(\bm{x}_i, y_i)\}$, in which $\bm{x}_i\in[0,1]^d$ is an image and $y_i\in\{1, ..., C\}$ is the ground-truth label, the adversary first modifies a proportion of training samples and then trains a model on the poisoned dataset. In particular, the adversary can insert a specific trigger (\eg, a patch) into a clean image $\bm{x}$ using a generic form~\cite{wang2019neural} as
\begin{equation}\label{eq:trigger}
    \bm{x}' \equiv \mathcal{A}(\bm{x}, \bm{m}, \bm{p}) = (\bm{1}-\bm{m})\cdot\bm{x} + \bm{m} \cdot \bm{p},
\end{equation}
where $\mathcal{A}$ is the function to apply the trigger, $\bm{m}\in\{0,1\}^d$ is the binary \emph{mask} to decide the position of the trigger, and $\bm{p}\in[0,1]^d$ is the trigger \emph{pattern}. The adversary takes a subset $\mathcal{D}'\subset\mathcal{D}$ containing $r\%$ of the training samples and creates poisoned data $\mathcal{D}_p'=\{(\bm{x}_i',y_i')|\bm{x}_i'=\mathcal{A}(\bm{x}_i, \bm{m}, \bm{p}), y_i'=y^t, (\bm{x}_i, y_i)\in\mathcal{D}'\}$, where $y^t$ is the adversary-specified target class. Finally, a classification model $f(\bm{x})$ is trained on the poisoned training dataset $(\mathcal{D}\setminus\mathcal{D}')\cup\mathcal{D}_p'$. The backdoor attack is considered successful if the model can classify the triggered images as the target class with a high success rate, while its accuracy on clean testing images is on a par with the normal model.
Although we introduce the simplest and most studied setting, our method can also be used under various threat models with experimental supports (Sec.~\ref{sec:4-4}).

\textbf{Defender:} We consider a more realistic black-box setting for the defender, in which the poisoned training dataset and the white-box model cannot be accessed. The defender can only query the trained model $f(\bm{x})$ as an oracle to obtain its predictions, but cannot acquire its gradients. We assume that $f(\bm{x})$ outputs predicted probabilities over all $C$ classes.
The goal of the defender is to distinguish whether $f(\bm{x})$ is normal or backdoored given a set of clean validation images or using synthetic samples in the case that the clean images are unavailable.

\subsection{Problem Formulation}\label{sec:3-2}

As discussed in~\cite{wang2019neural}, a model is regarded as backdoored if it requires much smaller modifications to cause misclassification to the target class than other uninfected ones. The reason is that the adversary usually wants to make the backdoor trigger inconspicuous. Thus, the defender can detect a backdoored model by judging whether any class needs significantly smaller modifications for misclassification.

Since the defender has no knowledge of the trigger pattern $(\bm{m}$, $\bm{p})$ and the true target class $y^t$, the potential trigger for each class $c$ can be reverse-engineered~\cite{wang2019neural} by solving 
\begin{equation}\label{eq:problem}\vspace{-0.4ex}
	\min _{\bm{m}, \bm{p}} \sum_{\bm{x}_i\in\bm{X}}\left\{\ell \big(c, f(A(\bm{x}_i, \bm{m}, \bm{p}))\big)+\lambda \cdot|\bm{m}|\right\},
\end{equation}
where $\bm{X}$ is the set of clean images to solve the optimization problem, $\ell(\cdot,\cdot)$ is the cross-entropy loss, and $\lambda$ is the balancing parameter. The optimization problem~\eqref{eq:problem} seeks to simultaneously generate a trigger $(\bm{m}, \bm{p})$ that leads to misclassification of clean images to the target class $c$ and minimize the trigger size measured by the $L_1$ norm of $\bm{m}$\footnote{Most of the previous backdoor attacks adopt a small patch as the backdoor trigger. Thus, the $L_1$ norm is an appropriate measure of trigger size.}. 
\emph{Neural Cleanse} (NC)~\cite{wang2019neural} relaxes the binary mask $\bm{m}$ to be continuous in $[0,1]^d$ and solves the problem~\eqref{eq:problem} by Adam~\cite{Kingma2014} with $\lambda$ tuned dynamically to ensure that more than $99\%$ clean images can be misclassified.
The optimization problem~\eqref{eq:problem} is solved for each class $c\in\{1, ..., C\}$ sequentially.

After obtaining the reversed triggers for all classes, we can identify whether the model has been backdoored based on outlier detection methods, which regard a class to be an infected one if the optimized mask $\bm{m}$ has much smaller $L_1$ norm. If all classes induce similar $L_1$ norm of the masks, the model is regarded to be normal.
The Median Absolute Deviation (MAD) is adopted in NC.
Although recent methods belonging to this defense category~\cite{chen2019deepinspect,guo2019tabor,harikumar2020scalable,qiao2019defending} have been proposed for better trigger restoration and outlier detection, all of these methods need access to model gradients for optimizing the triggers. In contrast, we propose an innovative method to solve the optimization problem~\eqref{eq:problem}, which can operate in the black-box manner without gradients.

\subsection{Black-box Backdoor Detection (B3D)}\label{sec:3-3}

We let $\mathcal{F}(\bm{m},\bm{p};c)$ denote the loss function in Eq.~\eqref{eq:problem} for notation simplicity. Under the black-box setting, the goal is to minimize $\mathcal{F}(\bm{m},\bm{p};c)$ without accessing model gradients. By sending queries to the trained model $f(\bm{x})$ and receiving its predictions, we can only obtain the value of $\mathcal{F}(\bm{m},\bm{p};c)$. 
Our proposed algorithm is motivated by \emph{Natural Evolution Strategies} (NES)~\cite{JMLR:v15:wierstra14a}, an effective gradient-free optimization method.
Similar to NES, the key idea of our algorithm is to learn a search distribution by using an estimated gradient on its parameters towards better loss value of interest.
But differently, we do not adopt natural gradients\footnote{We explain why we do not adopt natural gradients in Appendix~\ref{app:a}.} and the optimization involves a mixture of discrete and continuous variables (\ie, $\bm{m}$ and $\bm{p}$), which is known hard to solve~\cite{halstrup2016black}. To address this problem, we propose to utilize a discrete distribution to model $\bm{m}$ along with a continuous one to model $\bm{p}$, leading to a novel algorithm for optimization.

In particular, instead of minimizing $\mathcal{F}(\bm{m},\bm{p};c)$, we minimize the expected loss under the search distribution as
\begin{equation}\label{eq:nes}
    \min_{\bm{\theta}_{m},\bm{\theta}_{p}} \mathcal{J}(\bm{\theta}_m, \bm{\theta}_p) = \mathbb{E}_{\pi(\bm{m},\bm{p}|\bm{\theta}_m, \bm{\theta}_p)}[\mathcal{F}(\bm{m},\bm{p};c)],
\end{equation}
where $\pi(\bm{m},\bm{p}|\bm{\theta}_m, \bm{\theta}_p)$ is a distribution with parameters $\bm{\theta}_m$ and $\bm{\theta}_p$.
To define a proper distribution $\pi$ over $\bm{m}\in\{0,1\}^d$ and $\bm{p}\in[0,1]^d$, we let $g(\cdot)=\frac{1}{2}(\tanh(\cdot)+1)$ denote a normalization function and take the transformation of variable approach as
\begin{equation}\label{eq:variable}
\begin{gathered}
    \bm{m} \sim \mathrm{Bern}(g(\bm{\theta}_m)); \quad
    \bm{p} = g(\bm{p}'), \; \bm{p}' \sim \mathcal{N}(\bm{\theta}_p, \sigma^2),
\end{gathered}
\end{equation}
where $\bm{\theta}_m, \bm{\theta}_p \in \mathbb{R}^d$, $\mathrm{Bern}(\cdot)$ is the Bernoulli distribution, and $\mathcal{N}(\cdot, \cdot)$ is the Gaussian distribution with $\sigma$ being its standard deviation.
By adopting the formulation in Eq.~\eqref{eq:variable}, the constraints on $\bm{m}$ and $\bm{p}$ are satisfied while the optimization variables $\bm{\theta}_m$ and $\bm{\theta}_p$ are unconstrained.
Therefore, we do not need to relax $\bm{m}$ to be continuous in $[0,1]^d$ as the previous methods~\cite{guo2019tabor,wang2019neural} do and can perform optimization in the discrete domain. The experiments also reveal different behaviors between our method and baselines.

To solve the optimization problem~\eqref{eq:nes}, we need to estimate its gradients.
Note that $\bm{m}$ and $\bm{p}$ are independent, thus we can represent their joint distribution
$\pi(\bm{m},\bm{p}|\bm{\theta}_m, \bm{\theta}_p)$ by $\pi_1(\bm{m}|\bm{\theta}_m)\pi_2(\bm{p}|\bm{\theta}_p)$, in which $\pi_1(\bm{m}|\bm{\theta}_m)$ denotes the Bernoulli distribution of $\bm{m}$ and $\pi_2(\bm{p}|\bm{\theta}_p)$ denotes the transformation of Gaussian of $\bm{p}$, as defined in Eq.~\eqref{eq:variable}.
Hence, we can estimate the gradients of $\mathcal{J}(\bm{\theta}_m, \bm{\theta}_p)$ with respect to $\bm{\theta}_m$ and $\bm{\theta}_p$ separately.
To calculate $\nabla_{\bm{\theta}_m}\mathcal{J}(\bm{\theta}_m, \bm{\theta}_p)$, we denote $\mathcal{F}_1(\bm{m}) = \mathbb{E}_{\pi_2(\bm{p}|\bm{\theta}_p)}[\mathcal{F}(\bm{m},\bm{p};c)]$. Then we have
\begin{equation*}\label{eq:grad-m}
\begin{split}
    \nabla_{\bm{\theta}_m}\mathcal{J}(\bm{\theta}_m, \bm{\theta}_p) & = \nabla_{\bm{\theta}_m}\mathbb{E}_{\pi(\bm{m},\bm{p}|\bm{\theta}_m, \bm{\theta}_p)}[\mathcal{F}(\bm{m},\bm{p};c)] \\
    & = \nabla_{\bm{\theta}_m}\mathbb{E}_{\pi_1(\bm{m}|\bm{\theta}_m)}[\mathcal{F}_1(\bm{m})]\\
    & = \mathbb{E}_{\pi_1(\bm{m}|\bm{\theta}_m)} [\mathcal{F}_1(\bm{m}) \nabla_{\bm{\theta}_m} \log \pi_1(\bm{m}|\bm{\theta}_m)] \\
    & = \mathbb{E}_{\pi_1(\bm{m}|\bm{\theta}_m)} \big[\mathcal{F}_1(\bm{m})\cdot2(\bm{m} - g(\bm{\theta}_m)) \big].
\end{split}
\end{equation*}
In practice, we can obtain the estimate of the search gradient by approximating the expectation over $\bm{m}$ with $k$ samples $\bm{m}_1, ..., \bm{m}_k \sim \pi_1(\bm{m}|\bm{\theta}_m)$. There is also an expectation in $\mathcal{F}_1(\bm{m})$. We approximate it as $\mathcal{F}_1(\bm{m})\approx\mathcal{F}(\bm{m}, g(\bm{\theta}_p); c)$. Therefore, the gradient $\nabla_{\bm{\theta}_m}\mathcal{J}(\bm{\theta}_m, \bm{\theta}_p)$ can be obtained by
\vspace{-1ex}
\begin{equation}\label{eq:approx-m}\small
\begin{split}
    \nabla_{\bm{\theta}_m}\mathcal{J}(\bm{\theta}_m, \bm{\theta}_p)  \approx \frac{1}{k}\sum_{j=1}^k \mathcal{F}_1(\bm{m}_j)\cdot&2(\bm{m}_j - g(\bm{\theta}_m)) \\ \approx \frac{1}{k}\sum_{j=1}^k \mathcal{F}(\bm{m}_j, g(\bm{\theta}_p); c)\cdot&2(\bm{m}_j - g(\bm{\theta}_m)).
\end{split}
\end{equation}
As can be seen from Eq.~\eqref{eq:approx-m}, the gradient can be estimated by evaluating the loss function with random samples, which can be realized under the black-box setting through queries.

\begin{algorithm}[!t]
\small
\caption{Black-box backdoor detection (B3D)}\label{algo1}
\begin{algorithmic}[1]
\Require {A set of clean images $\bm{X}$; a target class $c$; the loss function in Eq.~\eqref{eq:problem} denoted as $\mathcal{F}(\bm{m},\bm{p};c)$; the search distribution $\pi$ defined in Eq.~\eqref{eq:variable}; standard deviation of Gaussian $\sigma$; the number of samples $k$; the number of iterations $T$.}
\Ensure {The parameters $\bm{\theta}_m$ and $\bm{\theta}_p$ of the search distribution $\pi$.}
\State Initialize $\bm{\theta}_m$ and $\bm{\theta}_p$;
\For {$t = 1$ to $T$}
\State $\hat{\bm{g}}_m \leftarrow \mathbf{0}$, $\hat{\bm{g}}_p \leftarrow \mathbf{0};$
\State Randomly draw a minibatch $\bm{X}_t$ from $\bm{X}$;
\For {$j=1$ to $k$}  \Comment{Estimate the gradient for $\bm{\theta}_m$}
\State Draw $\bm{m}_j \sim \mathrm{Bern}(g(\bm{\theta}_m))$;
\State $\hat{\bm{g}}_m \leftarrow \hat{\bm{g}}_m + \mathcal{F}(\bm{m}_j, g(\bm{\theta}_p);c)\cdot2(\bm{m}_j-g(\bm{\theta}_m))$;
\EndFor
\For {$j=1$ to $k$}  \Comment{Estimate the gradient for $\bm{\theta}_p$}
\State Draw $\bm{\epsilon}_j \sim \mathcal{N}(\mathbf{0}, \mathbf{I})$;
\State $\hat{\bm{g}}_p \leftarrow \hat{\bm{g}}_p + \mathcal{F}(g(\bm{\theta}_m), g(\bm{\theta}_p+\sigma\bm{\epsilon}_j); c)\cdot\bm{\epsilon}_j$;
\EndFor
\State Update $\bm{\theta}_m$ by $\bm{\theta}_m \leftarrow \mathrm{Adam.step}(\bm{\theta}_m, \frac{1}{k}\hat{\bm{g}}_m)$;
\State Update $\bm{\theta}_p$ by $\bm{\theta}_p \leftarrow \mathrm{Adam.step}(\bm{\theta}_p, \frac{1}{k\sigma}\hat{\bm{g}}_p)$;
\EndFor
\end{algorithmic}
\end{algorithm}

Similarly, we calculate the gradient $\nabla_{\bm{\theta}_p}\mathcal{J}(\bm{\theta}_m, \bm{\theta}_p)$ as
\begin{equation*}
\begin{split}
 \nabla_{\bm{\theta}_p}\mathcal{J}(\bm{\theta}_m, \bm{\theta}_p)
    & = \nabla_{\bm{\theta}_p}\mathbb{E}_{\pi_2(\bm{p}|\bm{\theta}_p)}[\mathcal{F}_2(\bm{p})] \\
    & = \mathbb{E}_{\bm{\epsilon}\sim\mathcal{N}(\mathbf{0},\mathbf{I})}\left[\mathcal{F}_2(g(\bm{\theta}_p+\sigma\bm{\epsilon}))\cdot\frac{\bm{\epsilon}}{\sigma}\right],
\end{split}
\end{equation*}
where $\mathcal{F}_2(\bm{p}) = \mathbb{E}_{\pi_1(\bm{m}|\bm{\theta}_m)}[\mathcal{F}(\bm{m},\bm{p};c)]$. We reparameterize $\bm{p}$ by $\bm{p}=g(\bm{p}')=g(\bm{\theta}_p+\sigma\bm{\epsilon})$, where $\bm{\epsilon}$ follows the standard Gaussian distribution $\mathcal{N}(\mathbf{0},\mathbf{I})$ to make the expression clearer. We approximate $\mathcal{F}_2(\bm{p})$ by $\mathcal{F}(g(\bm{\theta}_m), \bm{p};c)$ and obtain the estimate of the gradient $\nabla_{\bm{\theta}_p}\mathcal{J}(\bm{\theta}_m, \bm{\theta}_p)$ with another $k$ samples $\bm{\epsilon}_1, ..., \bm{\epsilon}_k \sim \mathcal{N}(\mathbf{0},\mathbf{I})$ as
\vspace{-1ex}
\begin{equation}\small\label{eq:grad-p}
\begin{split}
    \nabla_{\bm{\theta}_p}\mathcal{J}(\bm{\theta}_m, \bm{\theta}_p) \approx \frac{1}{k\sigma}\sum_{j=1}^k \mathcal{F}_2(g(\bm{\theta}_p+\sigma\bm{\epsilon}_j))&\cdot\bm{\epsilon}_j \\\vspace{-1ex}
    \approx \frac{1}{k\sigma}\sum_{j=1}^k \mathcal{F}(g(\bm{\theta}_m), g(\bm{\theta}_p+\sigma\bm{\epsilon}_j); c)&\cdot\bm{\epsilon}_j.
\end{split}
\end{equation}

After obtaining the estimated gradients, we can perform gradient descent to iteratively update the search distribution parameters $\bm{\theta}_m$ and $\bm{\theta}_p$. We adopt the same strategy as NC, that the Adam optimizer is used and the hyperparameter $\lambda$ in Eq.~\eqref{eq:problem} is adaptively tuned. We outline the proposed B3D algorithm in Algorithm~\ref{algo1}. In Step 4, we draw a minibatch $\bm{X}_t$ from the set of clean images $\bm{X}$ and evaluate the loss function $\mathcal{F}$ based on $\bm{X}_t$. Similar to NC, after we get the reversed triggers for every class $c$, we identify outliers based on the $L_1$ norm of the masks, and thereafter detect the backdoored model if any mask exhibits much smaller $L_1$ norm.
The details of our adopted outlier detection method will be introduced in the experiments.





\subsection{B3D with Synthetic Samples (B3D-SS)}\label{sec:3-4}

One limitation of the B3D algorithm as well as the previous methods~\cite{guo2019tabor,wang2019neural} is the dependence on a set of clean images, which could be unavailable in practice.
To perform backdoor detection in the absence of any clean data, a simple approach is to adopt a set of synthetic samples.
A good set of synthetic samples should satisfy that they are misclassified as the target class by adding the true trigger such that the true trigger is a solution of Eq.~\eqref{eq:problem} and there should not exist many solutions of Eq.~\eqref{eq:problem} such that we can recover the true trigger instead of obtaining other incorrect ones.

In practice, the synthetic samples could be drawn from a random distribution or created by generative models based on different datasets. 
Besides, we need to make these samples well-distributed over all classes when classified by the model $f(\bm{x})$ because in an extreme case that they are mostly classified as one class $c$, our algorithm would always generate a very small trigger for class $c$ based on the problem formulation~\eqref{eq:problem} no matter whether $c$ is the target class or not. 
To this end, we draw $n$ random images $\bm{X}^c:=\{\bm{x}_i^c\}_{i=1}^n$ for each class $c$ and minimize $\ell(c,f(\bm{x}_i^c))$ with respect to each image $\bm{x}_i^c$, in which $\ell(\cdot,\cdot)$ is the cross-entropy loss. 
Therefore, the resultant synthetic image $\bm{x}_i^c$ will be classified as $c$ by $f(\bm{x})$. 
Under the black-box setting, we use a gradient-free algorithm similar to Eq.~\eqref{eq:grad-p} to optimize $\bm{x}_i^c$ as
\vspace{-0.1ex}
\begin{equation}
    \bm{x}_i^c \leftarrow \bm{x}_i^c - \eta\cdot \frac{1}{k\sigma}\sum_{j=1}^k \ell(c, f(\bm{x}_i^c + \bm{\delta}_j))\cdot\bm{\delta}_j,
\end{equation}
where $\eta$ is the learning rate and $\bm{\delta}_1, ..., \bm{\delta}_k$ are drawn from $\mathcal{N}(\mathbf{0}, \mathbf{I})$.
The synthetic dataset is composed of the resultant images for all classes as $\bm{X} = \bigcup_{c=1}^{C}\bm{X}^c$, which is further used for reverse-engineering the trigger by Algorithm~\ref{algo1}.

\section{Experiments}\label{sec:4}
\textbf{Datasets.} We use CIFAR-10~\cite{krizhevsky2009learning}, German Traffic Sign Recognition Benchmark (GTSRB)~\cite{stallkamp2012man}, and ImageNet~\cite{russakovsky2015imagenet} datasets to conduct experiments.
On each dataset, we train hundreds of models to perform comprehensive evaluations. Some of them are normally trained while the others have been embedded backdoors.
We will detail the training and backdoor attack settings in Sec.~\ref{sec:cifar} for CIFAR-10, Sec.~\ref{sec:gtsrb} for GTSRB, and Sec.~\ref{sec:imagenet} for ImageNet.
Sec.~\ref{sec:4-4} shows the effectiveness of our methods under various settings.

\begin{table}[t]
    \begin{center}\footnotesize
    \begin{tabular}{l|p{10ex}<{\centering}|p{10ex}<{\centering}|p{10ex}<{\centering}}
    \hline
      & CIFAR-10 & GTSRB & ImageNet \\
     \hline\hline
     NC~\cite{wang2019neural} & 95.0\% & \bf100.0\% & \bf96.0\% \\ 
     TABOR~\cite{guo2019tabor} & 95.5\% & \bf100.0\% & 95.0\% \\
     B3D (Ours) & \bf97.5\% & \bf100.0\% & \bf96.0\% \\ 
     B3D-SS (Ours) & \bf97.5\% & \bf100.0\% & 95.5\% \\
    \hline
    \end{tabular}
    \end{center}
    \vspace{-3ex}
    \caption{The backdoor detection accuracy of NC, TABOR, B3D, and B3D-SS on the CIFAR-10, GTSRB, and ImageNet datasets.}
    \vspace{-1ex}
    \label{tab:overall}
\end{table}

\textbf{Compared methods.} We compare B3D and B3D-SS with Neural Cleanse (NC)~\cite{wang2019neural} and TABOR~\cite{guo2019tabor}, which are typical and state-of-the-art methods based on model gradients.
In B3D and B3D-SS, we set the number of samples $k$ as $50$, the standard deviation of Gaussian $\sigma$ as $0.1$, the learning rate of the Adam optimizer as $0.05$. We provide the implementation details and more analyses on the hyperparameters in Appendix~\ref{app:b}. The optimization is conducted until convergence.
After obtaining the distribution parameters $\bm{\theta}_m$ and $\bm{\theta}_p$, we could generate the mask by discretization as $\bm{m}=\mathbf{1}[g(\bm{\theta}_m) \geq 0.5]$ and the pattern as $\bm{p}=g(\bm{\theta}_p)$.
To compare with the baselines, we adopt the ``soft'' mask $g(\bm{\theta}_m)$ in experiments. TABOR introduces several regularizations to improve the performance of backdoor detection. Although our algorithm is based on the problem formulation~\eqref{eq:problem} similar to NC, it can easily be extended to others (\eg, TABOR), which we leave to future work.

\textbf{Outlier detection.} Given the reversed triggers for all classes, we calculate their $L_1$ norm and perform outlier detection to identify very small triggers (\ie, outliers).
We observe that the Median Absolute Deviation (MAD) adopted in NC performs poorly in some cases due to the assumption of a Gaussian distribution, which does not hold for all cases, especially when the number of classes $C$ is small.
Hence, we further add a heuristic rule to identify small triggers by judging whether the $L_1$ norm of any mask is smaller than one fourth of their median. This method is also applied to NC to improve the baseline performance.

\begin{table*}[!t]
\footnotesize
\begin{center}
\begin{tabular}{c|c|c|c|cc|cccc}

\hline
\multirow{2}{*}{Model} & \multirow{2}{*}{Accuracy} & \multirow{2}{*}{ASR} & \multirow{2}{*}{Method} & \multicolumn{2}{c|}{Reversed Trigger} & \multicolumn{4}{c}{Detection Results}\\
\cline{5-10}
& & & & $L_1$ norm & ASR & Case I & Case II & Case III & Case IV \\
\hline\hline
Normal & 89.30\% & N/A & \tabincell{l}{NC~\cite{wang2019neural} \\ TABOR~\cite{guo2019tabor} \\ B3D (Ours) \\ B3D-SS (Ours)} & \tabincell{c}{N/A \\ N/A \\ N/A \\ N/A} & \tabincell{c}{N/A \\ N/A \\ N/A \\ N/A} & \tabincell{c}{N/A \\ N/A \\ N/A \\ N/A} & \tabincell{c}{N/A \\ N/A \\ N/A \\ N/A} & \tabincell{c}{8/50 \\ 4/50 \\ 2/50 \\ 3/50} & \tabincell{c}{42/50 \\ 46/50 \\ 48/50 \\ 47/50 } \\
\hline
\tabincell{c}{Backdoored \\ ($1\times1$ trigger)} & 88.35\% & 99.75\% & \tabincell{l}{NC~\cite{wang2019neural} \\ TABOR~\cite{guo2019tabor} \\ B3D (Ours) \\ B3D-SS (Ours)} & \tabincell{c}{0.588 \\ 0.672 \\ 0.820 \\ 3.734} & \tabincell{c}{98.76\% \\ 99.11\% \\ 99.29\% \\ 99.98\%} & \tabincell{c}{40/50 \\ 36/50 \\ 36/50 \\ 35/50} & \tabincell{c}{9/50 \\ 13/50 \\ 12/50 \\ 15/50} & \tabincell{c}{0/50 \\ 0/50 \\ 0/50 \\ 0/50} & \tabincell{c}{1/50 \\ 1/50 \\ 2/50 \\ 0/50} \\
\hline
\tabincell{c}{Backdoored \\ ($2\times2$ trigger)} & 88.51\% & 100.00\% & \tabincell{l}{NC~\cite{wang2019neural} \\ TABOR~\cite{guo2019tabor} \\ B3D (Ours) \\ B3D-SS (Ours)} & \tabincell{c}{1.508 \\ 2.256 \\ 2.310 \\ 2.867} & \tabincell{c}{98.81\% \\ 99.21\% \\ 98.94\% \\ 99.13\%} & \tabincell{c}{47/50 \\ 44/50 \\ 47/50 \\ 47/50} & \tabincell{c}{2/50 \\ 3/50 \\ 3/50 \\ 2/50} & \tabincell{c}{0/50 \\ 0/50 \\ 0/50 \\ 0/50} & \tabincell{c}{1/50 \\ 3/50 \\ 0/50 \\ 1/50} \\
\hline
\tabincell{c}{Backdoored \\ ($3\times3$ trigger)}& 88.57\% & 100.00\% & \tabincell{l}{NC~\cite{wang2019neural} \\ TABOR~\cite{guo2019tabor} \\ B3D (Ours) \\ B3D-SS (Ours)} & \tabincell{c}{2.264 \\ 2.493 \\ 3.521 \\ 3.856} & \tabincell{c}{98.71\% \\ 98.84\% \\ 98.87\% \\ 96.97\%} & \tabincell{c}{49/50 \\ 48/50 \\ 47/50 \\ 47/50} & \tabincell{c}{1/50 \\ 1/50 \\ 2/50 \\ 2/50} & \tabincell{c}{0/50 \\ 0/50 \\ 0/50 \\ 0/50} & \tabincell{c}{0/50 \\ 1/50 \\ 1/50 \\ 1/50} \\
\hline
\end{tabular}
\end{center}
\vspace{-4.5ex}
\caption{The results of backdoor detection on CIFAR-10. For normal and backdoored models with different trigger sizes, we show their average accuracy and backdoor attack success rates (ASR). For the four backdoor detection methods --- NC, TABOR, B3D, and B3D-SS, we report the $L_1$ norm and attack success rates of the reversed trigger corresponding to the target class, as well as the detection results in four cases.}
\label{tab:cifar}
\vspace{-1ex}
\end{table*}

\textbf{Evaluations.} Table~\ref{tab:overall} shows the overall backdoor detection accuracy of all methods on three datasets. Our methods achieve comparable or even better performance than the baselines, while rely on weak assumptions (\ie, black-box setting) for backdoor detection, validating the effectiveness of our methods.
In addition to the coarse results, we further conduct sophisticated analyses of the performance of different methods on each dataset. Specifically, we consider four cases of backdoor detection for an algorithm $\mathcal{A}$:
\begin{itemize}\vspace{-1.3ex}
    \item \textbf{Case I}: $\mathcal{A}$ successfully identifies a backdoored model and correctly discovers the true target class without reporting other backdoor attacks for uninfected classes.\vspace{-1.3ex}
    \item \textbf{Case II}: $\mathcal{A}$ successfully identifies a backdoored model but discovers multiple backdoor attacks for both the true target class and other uninfected classes.\vspace{-1.3ex}
    \item \textbf{Case III}: $\mathcal{A}$ wrongly identifies a normal model as backdoored or wrongly discovers backdoor attacks for uninfected classes excluding the true target class of a backdoored model.\vspace{-1.3ex}
    \item \textbf{Case IV}: $\mathcal{A}$ successfully identifies a normal model or wrongly identifies a backdoored model as normal.\vspace{-1.3ex}
\end{itemize}

In the following, we introduce the detailed experimental results on each dataset.

\subsection{CIFAR-10}\label{sec:cifar}
We adopt the ResNet-18~\cite{He2015} architecture on CIFAR-10. The backdoor attacks are implemented using the BadNets approach~\cite{gu2017badnets}. We consider the triggers of size $1\times1$, $2\times2$, and $3\times3$. For each size, we train $50$ backdoored models using different triggers and target classes with $5$ models per target class. The triggers are generated in random positions and have random colors. 
We poison $10\%$ training data. Besides, we also train $50$ normal models with different random seeds, resulting in a total number of $200$ models. We train them for $200$ epochs without using data augmentation. The accuracy on the clean test set and the backdoor attack success rates (ASR) are shown in Table~\ref{tab:cifar} (column 2-3).

\begin{figure}[t]
\centering
\includegraphics[width=0.96\linewidth]{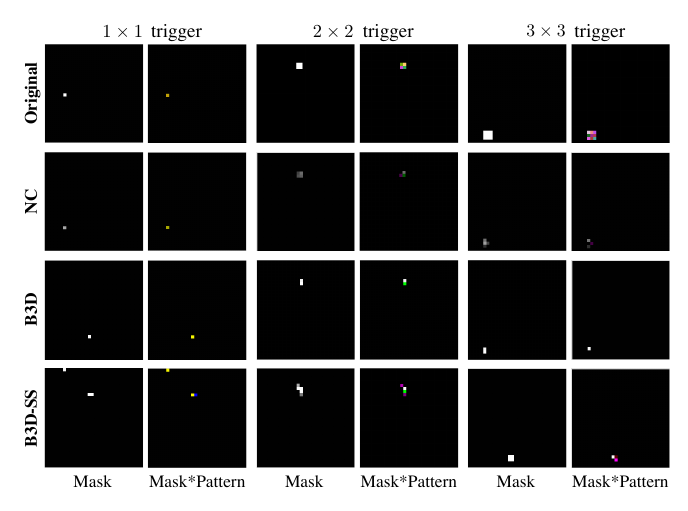}
\vspace{-1.5ex}
\caption{Visualization of the original triggers and the reversed triggers optimized by NC, B3D, and B3D-SS on CIFAR-10.}
\label{fig:vis}
\vspace{-1ex}
\end{figure}
To perform backdoor detection, NC, TABOR, and B3D adopt the $10,000$ clean test images, while B3D-SS adopts $1,000$ synthetic images with $100$ per class. In Table~\ref{tab:cifar}, we report the $L_1$ norm and the attack success rates (ASR) of the reversed trigger corresponding to the true target class for the backdoored models. We also report the number of models belonging to the four cases of backdoor detection. In Fig.~\ref{fig:vis}, we visualize the original triggers and the reversed triggers optimized by NC, B3D, and B3D-SS with different trigger sizes. 
From the results, we draw the findings below.
\begin{table*}[!t]
\footnotesize

\begin{center}
\begin{tabular}{c|c|c|c|cc|cccc}

\hline
\multirow{2}{*}{Model} & \multirow{2}{*}{Accuracy} & \multirow{2}{*}{ASR} & \multirow{2}{*}{Method} & \multicolumn{2}{c|}{Reversed Trigger} & \multicolumn{4}{c}{Detection Results}\\
\cline{5-10}
& & & & $L_1$ norm & ASR & Case I & Case II & Case III & Case IV \\
\hline\hline
Normal & 98.84\% & N/A & \tabincell{l}{NC~\cite{wang2019neural} \\ TABOR~\cite{guo2019tabor} \\ B3D (Ours) \\ B3D-SS (Ours)} & \tabincell{c}{N/A \\ N/A \\ N/A \\ N/A} & \tabincell{c}{N/A \\ N/A \\ N/A \\ N/A} & \tabincell{c}{N/A \\ N/A \\ N/A \\ N/A} & \tabincell{c}{N/A \\ N/A \\ N/A \\ N/A} & \tabincell{c}{0/43 \\ 0/43 \\ 0/43 \\ 0/43} & \tabincell{c}{43/43 \\ 43/43 \\ 43/43 \\ 43/43 } \\
\hline
\tabincell{c}{Backdoored \\ ($1\times1$ trigger)} & 98.74\% & 99.53\% & \tabincell{l}{NC~\cite{wang2019neural} \\ TABOR~\cite{guo2019tabor} \\ B3D (Ours) \\ B3D-SS (Ours)} & \tabincell{c}{0.737 \\ 0.543 \\ 0.922 \\ 3.079} & \tabincell{c}{98.90\% \\ 99.24\% \\ 98.86\% \\ 100.00\%} & \tabincell{c}{14/43 \\ 19/43 \\ 10/43 \\ 12/43} & \tabincell{c}{29/43 \\ 24/43 \\ 33/43 \\ 31/43} & \tabincell{c}{0/43 \\ 0/43 \\ 0/43 \\ 0/43} & \tabincell{c}{0/43 \\ 0/43 \\ 0/43 \\ 0/43} \\
\hline
\tabincell{c}{Backdoored \\ ($2\times2$ trigger)} & 98.79\% & 100.00\% & \tabincell{l}{NC~\cite{wang2019neural} \\ TABOR~\cite{guo2019tabor} \\ B3D (Ours) \\ B3D-SS (Ours)} & \tabincell{c}{1.439 \\ 1.783 \\ 2.260 \\ 2.351} & \tabincell{c}{98.75\% \\ 99.15\% \\ 99.04\% \\ 97.96\%} & \tabincell{c}{27/43 \\ 22/43 \\ 27/43 \\ 25/43} & \tabincell{c}{16/43 \\ 21/43 \\ 16/43 \\ 18/43} & \tabincell{c}{0/43 \\ 0/43 \\ 0/43 \\ 0/43} & \tabincell{c}{0/43 \\ 0/43 \\ 0/43 \\ 0/43} \\
\hline
\tabincell{c}{Backdoored \\ ($3\times3$ trigger)}& 98.79\% & 100.00\% & \tabincell{l}{NC~\cite{wang2019neural} \\ TABOR~\cite{guo2019tabor} \\ B3D (Ours) \\ B3D-SS (Ours)} & \tabincell{c}{2.264 \\ 2.764 \\ 3.758 \\ 3.048} & \tabincell{c}{98.71\% \\ 99.22\% \\ 98.87\% \\ 94.87\%} & \tabincell{c}{39/43 \\ 35/43 \\ 34/43 \\ 33/43} & \tabincell{c}{4/43 \\ 8/43 \\ 9/43 \\ 10/43} & \tabincell{c}{0/43 \\ 0/43 \\ 0/43 \\ 0/43} & \tabincell{c}{0/43 \\ 0/43 \\ 0/43 \\ 0/43} \\
\hline
\end{tabular}
\end{center}
\vspace{-4.5ex}
\caption{The results of backdoor detection on GTSRB. For normal and backdoored models with different trigger sizes, we show their average accuracy and backdoor attack success rates (ASR). For the four backdoor detection methods --- NC, TABOR, B3D, and B3D-SS, we report the $L_1$ norm and attack success rates of the reversed trigger corresponding to the target class, as well as the detection results in four cases.}
\label{tab:gtsrb}
\vspace{-1ex}
\end{table*}

First, the reversed triggers of NC have smaller $L_1$ norm than B3D and B3D-SS. It is reasonable since NC performs direct optimization using gradients. However, as NC relaxes the mask $\bm{m}$ to be continuous in $[0,1]^d$, the optimized masks shown in Fig.~\ref{fig:vis} tend to have small amplitudes. For B3D and B3D-SS, since we let $\bm{m}$ follow the Bernoulli distribution, the optimized masks have values closer to $0$ (black) or $1$ (white), which is in accordance with the formulation~\eqref{eq:trigger}.

Second, as can be seen from Table~\ref{tab:cifar}, NC wrongly identifies more normal models as backdoored (\ie, $8$ out of $50$) than B3D and B3D-SS. It is also because that NC relaxes the mask $\bm{m}$ to $[0,1]^d$. Thus NC sometimes optimizes a mask with small $L_1$ norm for an uninfected class, which does not resemble true backdoor patterns and is identified as an outlier by MAD. But B3D and B3D-SS perform optimization in the discrete domain, which are less prone to this problem. We will further discuss this phenomenon in Appendix~\ref{app:c}.

\begin{figure}[t]
\centering
\includegraphics[width=0.82\linewidth]{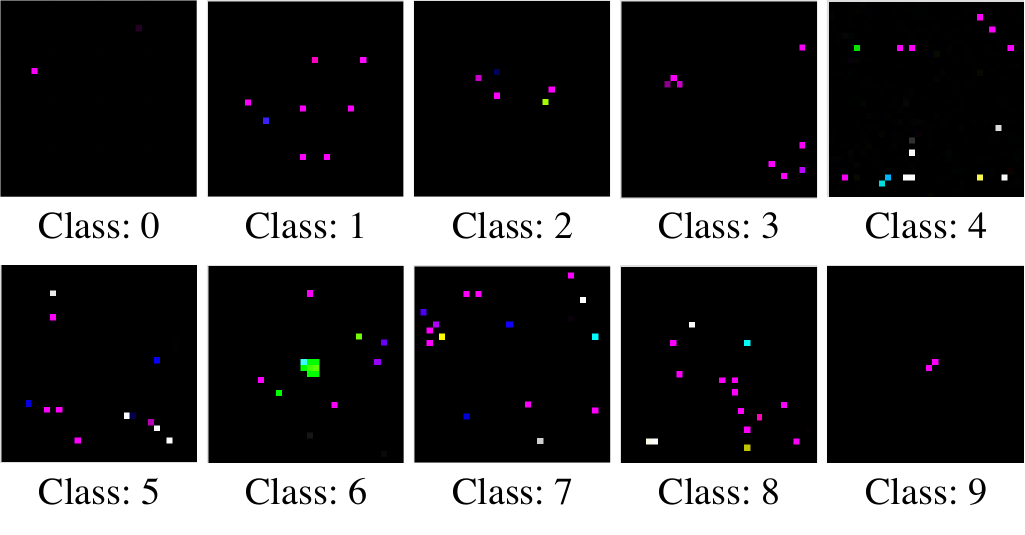}
\vspace{-1ex}
\caption{Visualization of the reversed triggers optimized by B3D for all classes on CIFAR-10. The true target class is 0, but B3D reports two backdoor attacks corresponding to class 0 and 9.}
\label{fig:vis2}
\vspace{-1ex}
\end{figure}

Third, we find that many backdoored models, especially those with $1\times1$ triggers, can be found multiple backdoors (\ie, Case II), as shown in Table~\ref{tab:cifar}. We verify that a chosen backdoored model truly has two backdoors in Fig.~\ref{fig:vis2}. So we think that backdoor attacks through data poisoning can not only affect the behavior of the model corresponding to the true target class, but also interfere other uninfected classes.

Fourth, as shown in Fig.~\ref{fig:vis}, the reversed triggers can have different positions and patterns compared with the original triggers. It indicates that a backdoored model would learn a distribution of triggers by generalizing the original one~\cite{qiao2019defending}. We provide further analysis on the effective input positions of backdoor attacks in Appendix~\ref{app:d}.

\subsection{GTSRB}\label{sec:gtsrb}

We adopt the same model architecture (\ie, ResNet-18) and backdoor injection method (\ie, BadNets) as in CIFAR-10. Since GTSRB has $43$ classes, we train one backdoored model for each class, resulting in $43$ backdoored models for a specific trigger size. We train another $43$ normal models for comparison. These models are trained for $50$ epochs. 
For backdoor inspection, NC, TABOR, and B3D adopt the $12,630$ clean test images for optimization, while B3D-SS generates $4,300$ synthetic images with $100$ per class.

The detailed experimental results on the statistics of the reversed triggers and the backdoor detection accuracy are presented in Table~\ref{tab:gtsrb}. The observations are consistent with those on CIFAR-10. We also find that the backdoor detection accuracy achieves $100\%$.
We think that the perfect detection accuracy is partially a consequence of more classes in this dataset, which enables the outlier detection method to correctly find outliers with more data points.

\subsection{ImageNet}\label{sec:imagenet}
\begin{table*}[!t]
\footnotesize
\begin{center}
\begin{tabular}{c|c|c|c|cc|cccc}

\hline
\multirow{2}{*}{Model} & \multirow{2}{*}{Accuracy} & \multirow{2}{*}{ASR} & \multirow{2}{*}{Method} & \multicolumn{2}{c|}{Reversed Trigger} & \multicolumn{4}{c}{Detection Results}\\
\cline{5-10}
& & & & $L_1$ norm & ASR & Case I & Case II & Case III & Case IV \\
\hline\hline
Normal & 88.46\% & N/A & \tabincell{l}{NC~\cite{wang2019neural} \\ TABOR~\cite{guo2019tabor} \\ B3D (Ours) \\ B3D-SS (Ours)} & \tabincell{c}{N/A \\ N/A \\ N/A \\ N/A} & \tabincell{c}{N/A \\ N/A \\ N/A \\ N/A} & \tabincell{c}{N/A \\ N/A \\ N/A \\ N/A} & \tabincell{c}{N/A \\ N/A \\ N/A \\ N/A} & \tabincell{c}{2/50 \\ 1/50 \\ 0/50 \\ 1/50} & \tabincell{c}{48/50 \\ 49/50 \\ 50/50 \\ 49/50 } \\
\hline
\tabincell{c}{Backdoored \\ (Trigger \parbox[c]{3ex}{
      \includegraphics[width=1\linewidth]{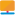}})} & 87.91\% & 99.95\% & \tabincell{l}{NC~\cite{wang2019neural} \\ TABOR~\cite{guo2019tabor} \\ B3D (Ours) \\ B3D-SS (Ours)} & \tabincell{c}{62.093 \\ 57.569 \\ 86.083 \\ 120.822} & \tabincell{c}{99.11\% \\ 99.25\% \\ 99.14\% \\ 97.57\%} & \tabincell{c}{45/50 \\ 43/50 \\ 43/50 \\ 42/50} & \tabincell{c}{0/50 \\ 0/50 \\ 0/50 \\ 0/50} & \tabincell{c}{0/50 \\ 0/50 \\ 0/50 \\ 0/50} & \tabincell{c}{5/50 \\ 7/50 \\ 7/50 \\ 8/50} \\
\hline
\tabincell{c}{Backdoored \\ (Trigger \parbox[c]{3ex}{
      \includegraphics[width=1\linewidth]{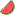}})} & 87.52\% & 99.68\% & \tabincell{l}{NC~\cite{wang2019neural} \\ TABOR~\cite{guo2019tabor} \\ B3D (Ours) \\ B3D-SS (Ours)} & \tabincell{c}{20.610 \\ 22.035 \\ 23.497 \\ 24.124} & \tabincell{c}{99.12\% \\ 99/24\% \\ 99.09\% \\ 97.15\%} & \tabincell{c}{50/50 \\ 47/50 \\ 50/50 \\ 44/50} & \tabincell{c}{0/50 \\ 2/50 \\ 0/50 \\ 6/50} & \tabincell{c}{0/50 \\ 0/50 \\ 0/50 \\ 0/50} & \tabincell{c}{0/50 \\ 1/50 \\ 0/50 \\ 0/50} \\
\hline
\tabincell{c}{Backdoored \\ (Trigger \parbox[c]{3ex}{
      \includegraphics[width=1\linewidth]{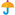}})} & 87.39\% & 99.94\% & \tabincell{l}{NC~\cite{wang2019neural} \\ TABOR~\cite{guo2019tabor} \\ B3D (Ours) \\ B3D-SS (Ours)} & \tabincell{c}{38.701 \\ 37.499 \\ 56.636 \\ 37.253} & \tabincell{c}{99.14\% \\ 99.20\% \\ 99.13\% \\ 97.44\%} & \tabincell{c}{48/50 \\ 46/50 \\ 48/50 \\ 49/50} & \tabincell{c}{1/50 \\ 3/50 \\ 1/50 \\ 1/50} & \tabincell{c}{0/50 \\ 0/50 \\ 0/50 \\ 0/50} & \tabincell{c}{1/50 \\ 1/50 \\ 1/50 \\ 0/50} \\
\hline
\end{tabular}
\end{center}
\vspace{-4.5ex}
\caption{The results of backdoor detection on ImageNet. For normal and backdoored models with different triggers, we show their average accuracy and backdoor attack success rates (ASR). For the dour backdoor detection methods --- NC, TABOR, B3D, and B3D-SS, we report the $L_1$ norm and attack success rates of the reversed trigger corresponding to the target class, as well as the detection results in four cases.}
\label{tab:imagenet}
\vspace{-1ex}
\end{table*}

Since the original ImageNet dataset contains more than $14$ million images, it is hard to train hundreds of models on it. Hence, we use a subset of $10$ classes, where each class has $\sim1,300$ images. The test set is composed of $500$ images with $50$ per class. These images have the resolution of $224\times224$.
We also adopt the ResNet-18 model. 
For backdoor attacks, we consider three pre-defined patterns shown in Table~\ref{tab:imagenet} of size $15\times15$ as the triggers rather than the randomly generated triggers.
Similar to the experimental settings on CIFAR-10, we train $50$ backdoored models using each trigger, in which $5$ models per target class are trained with the trigger stamped at random positions. 
For backdoor detection, NC, TABOR, and B3D adopt the $500$ test images, while B3D-SS utilizes $1,000$ synthetic images generated by BigGAN~\cite{brock2018large}, due to the poor performance of using random noises in the high-dimensional image space of ImageNet.

We show the backdoor detection results on ImageNet in Table~\ref{tab:imagenet}. Similar to the results on CIFAR-10 and GTSRB, our proposed B3D and B3D-SS methods can achieve comparable performance with the baselines. The reversed triggers also exhibit different visual appearance compared with the original triggers, as shown in Appendix~\ref{app:e}.

\subsection{Ablation Study on More Settings}\label{sec:4-4}
Besides the above experiments, we further demonstrate the generalizability of the proposed methods B3D and B3D-SS by considering more settings, including:
\begin{itemize}\vspace{-1.3ex}
    \item \textbf{Other backdoor attacks.} We study the blended injection attack~\cite{chen2017targeted} and label-consistent attack~\cite{turner2019label} to insert backdoors besides BadNets. 
    \vspace{-1.3ex}
    \item \textbf{Different model architectures.} We study a VGG~\cite{simonyan2014very} model architecture besides the ResNet model. \vspace{-1.3ex}
    \item \textbf{Data augmentation.} We investigate the effects of data augmentation for backdoor attacks and detection. \vspace{-1.3ex}
    \item \textbf{Multiple infected classes with different triggers.} We consider the scenario that multiple backdoors with different target classes are embedded in a model. \vspace{-1.3ex}
    \item \textbf{Single infected class with multiple triggers.} We consider the scenario that multiple backdoors with a single target class are embedded in a model. \vspace{-1ex}
\end{itemize}
Due to the space limitation, the complete experiments on these settings are deferred to Appendix~\ref{app:f}.

\section{Mitigation of Backdoor Attacks}\label{sec:mitigation}

Once a backdoor attack has been detected, we can further mitigate the backdoor to preserve the model utility for users. Under the studied black-box setting, we are unable to modify the model weights, such that the typical re-training or fine-tuning~\cite{liu2018fine,tran2018spectral,wang2019neural} strategies cannot be utilized. 
In this section, we introduce a simple and effective strategy for reliable predictions by rejecting any adversary-crafted input with the backdoor trigger stamped during inference.

\begin{table}[t]
    \begin{center}\footnotesize
    \begin{tabular}{l|p{10ex}<{\centering}|p{10ex}<{\centering}|p{10ex}<{\centering}}
    \hline
      & CIFAR-10 & GTSRB & ImageNet \\
     \hline\hline
     STRIP~\cite{gao2019strip} & 0.9332 & 0.4937 & 0.7126 \\
     Kernel Density~\cite{jin2020unified} & 0.9585 & 0.9874 & 0.9328 \\
     \hline
     NC~\cite{wang2019neural} & 0.9948 & \bf0.9962 & 0.9812 \\
     TABOR~\cite{guo2019tabor} & 0.9937 & 0.9953 & \bf0.9842 \\
     B3D (Ours) & \bf0.9958 & 0.9946 & 0.9806 \\
     B3D-SS (Ours) & 0.9856 & 0.9924 & 0.9833 \\
     \hline
    \end{tabular}
    \end{center}
    \vspace{-3ex}
    \caption{The AUC-scores of detecting triggered inputs during inference on the CIFAR-10, GTSRB, and ImageNet datasets. We use the metric $\mathcal{S}(\bm{x})$ in Eq.~\eqref{eq:metric} with the reversed triggers given by NC, TABOR, B3D, and B3D-SS, respectively. The performance is compared with additional baselines, including STRIP~\cite{gao2019strip} and the kernel density method~\cite{jin2020unified}.}
    \vspace{-1ex}
    \label{tab:mitigation}
\end{table}

Assume that we have detected a backdoored model $f(\bm{x})$ and discovered the true target class $y^t$. The optimized trigger for the target class is denoted as $(\bm{m}, \bm{p})$. The basic intuition behind our method is as follows. For a clean input $\bm{x}_c$ and a triggered input $\bm{x}_a$ crafted by the adversary, the predictions of $\bm{x}_c$ and $\mathcal{A}(\bm{x}_c, \bm{m}, \bm{p})$ by applying the reversed trigger are extremely different, while the predictions of $\bm{x}_a$ and $\mathcal{A}(\bm{x}_a, \bm{m}, \bm{p})$ are similar. 
The rationale is that both $\bm{x}_a$ and $\mathcal{A}(\bm{x}_a, \bm{m}, \bm{p})$ have the trigger stamped and are classified as the target class $y^t$ with similar probability distributions.
Therefore, for an arbitrary input $\bm{x}$, we let
\begin{equation}\label{eq:metric}
    \mathcal{S}(\bm{x}) = \mathcal{D}_{\mathrm{KL}}\big(f(\bm{x})||f(\mathcal{A}(\bm{x}, \bm{m}, \bm{p}))\big)
\end{equation}
measure the similarity between the model predictions $f(\bm{x})$ and $f(\mathcal{A}(\bm{x}, \bm{m}, \bm{p}))$, 
where $\mathcal{D}_{\mathrm{KL}}$ is the Kullback-Leibler divergence.
If $\mathcal{S}(\bm{x})$ is large, $\bm{x}$ is probably a clean input, and otherwise $\bm{x}$ has the trigger stamped, which will be rejected without a prediction.
Based on the metric $\mathcal{S}(\bm{x})$, we perform binary classification of clean inputs and triggered inputs on each dataset's test set. We report the AUC-scores averaged over all backdoored models in Table~\ref{tab:mitigation}. Using the reversed triggers optimized by any method, the proposed strategy can reliably detect the triggered inputs, achieving better performance than alternative baselines~\cite{gao2019strip,jin2020unified}.

\section{Conclusion}

In this paper, we proposed a black-box backdoor detection (B3D) method to identify backdoored models under the black-box setting. By formulating backdoor detection as an optimization problem, B3D solves the problem with model queries only. B3D can also be utilized with synthetic samples. We further introduced a simple and effective strategy to mitigate the discovered backdoor for reliable predictions.
We conducted extensive experiments on several datasets to demonstrate the effectiveness of the proposed methods. Our methods reach comparable or even better performance than the previous methods based on stronger assumptions.

{\small
\bibliographystyle{ieee_fullname}
\bibliography{egbib}
}

\clearpage
\appendix
\captionsetup{font={small}}

\section{Natural Gradients}\label{app:a}

Natural Evolution Strategies (NES)~\cite{JMLR:v15:wierstra14a} adopt the natural gradients for optimization, because \cite{JMLR:v15:wierstra14a} illustrates that the plain search gradients make the optimization very unstable when sampling from a Gaussian distribution with the learnable mean and covariance matrix.
The natural gradient is defined as
\begin{equation}
    \widetilde{\nabla}_{\bm{\theta}} \mathcal{J} = \mathbf{F}^{-1}\nabla_{\bm{\theta}} \mathcal{J}(\bm{\theta}),
\end{equation}
where $\bm{\theta}$ denotes the search distribution parameter and $\mathbf{F}$ is Fisher information matrix as
\begin{equation}
    \mathbf{F} = \mathbb{E}_{\pi(\cdot|\bm{\theta})}\left[\nabla_{\bm{\theta}} \log \pi(\cdot|\bm{\theta})\nabla_{\bm{\theta}} \log \pi(\cdot|\bm{\theta})^\top\right].
\end{equation}

In our problem, we could also calculate the Fisher information matrices for the search distributions $\pi_1(\bm{m}|\bm{\theta}_{m})$ and $\pi_2(\bm{p}|\bm{\theta}_{p})$. For $\pi_1(\bm{m}|\bm{\theta}_{m})$, we have
\begin{equation*}
\begin{split}
    \mathbf{F} & = \mathbb{E}_{\pi_1(\bm{m}|\bm{\theta}_m)}\left[\nabla_{\bm{\theta}_m} \log \pi_1(\bm{m}|\bm{\theta}_m)\nabla_{\bm{\theta}_m} \log \pi_1(\bm{m}|\bm{\theta}_m)^\top\right] \\
    & = \mathbb{E}_{\pi_1(\bm{m}|\bm{\theta}_m)}\left[4(\bm{m} - g(\bm{\theta}_m))(\bm{m} - g(\bm{\theta}_m))^\top\right] \\
    & = 4\cdot\mathrm{diag}\left(g(\bm{\theta}_m) (1 - g(\bm{\theta}_m))\right),
\end{split}
\end{equation*}
where $\mathrm{diag}(\cdot)$ denotes the diagonal matrix.
If the optimization on $\bm{\theta}_{m}$ is nearly converged, $g(\bm{\theta}_m)$ tends to be close to $0$ or $1$ since the mask $\bm{m}$ sampled from $\mathrm{Bern}(g(\bm{\theta}_m))$ should not change dramatically with different tries. Therefore, the diagonal elements in $\mathbf{F}$ tend to be $0$ and those of $\mathbf{F}^{-1}$ tend to be $+\infty$. Consequently, the optimization would be rather unstable if we adopt natural gradients.

For $\pi_2(\bm{p}|\bm{\theta}_{p})$, note that the variance of the Gaussian distribution is fixed, and thus the Fisher information matrix becomes $\mathbf{I}$. In this case, the natural gradients are the same as the plain gradients. Hence, we do not adopt natural gradients for optimization in our problem.

\section{Implementation Details and Hyperparameters}\label{app:b}

The implementation of Neural Cleanse (NC)~\cite{wang2019neural} is based on the official source code\footnote{\url{https://github.com/bolunwang/backdoor}.}. The source code of TABOR~\cite{guo2019tabor} was not released by the authors. Thus we implement TABOR based on another (unofficial) implementation\footnote{\url{https://github.com/UsmannK/TABOR}.}.
Our proposed B3D follows a similar optimization process to NC but replaces the white-box gradients by the estimated gradients, as detailed in Sec.~\ref{sec:3-3}. The hyperparameter $\lambda$ in Eq.~\eqref{eq:problem} is adjusted dynamically according to the backdoor attack success rate of several past optimization iterations, which is also based on the implementation of NC. 

In B3D and B3D-SS, we introduce one critical hyperparameter $k$ (\ie, the number of samples to estimate the gradient), which can affect the performance of backdoor detection. If $k$ is too small, the estimated gradient exhibits a large variance, making the optimization rather unstable. Otherwise, if $k$ is too large, the optimization needs more queries and time. Therefore, we need to choose a suitable $k$ to have a relatively small variance and make the optimization efficient. So we choose $k=50$ in the main experiments and we find that using $k\in[20,100]$ leads to similar results.
The optimization process is not very sensitive to different $k$.

In B3D-SS, we adopt a set of synthetic samples to perform optimization. The quality of the synthetic samples is also a critical factor to affect the performance of our algorithm. There are two important aspects --- the number of synthetic samples and the generation method of synthetic samples. Intuitively speaking, more synthetic samples are beneficial for reverse-engineering the true trigger since the optimization process would not easily drop into local minima. Empirically, we observe that using thousands of synthetic samples is sufficient for optimization, and thus we do not try to use more. On the other hand, the generation method of synthetic samples depends on the datasets. For CIFAR-10 and GTSRB, we find that using randomly generated samples from a uniform distribution can help to restore the true trigger. But for ImageNet, the randomly generated samples are not helpful since the input dimension is much higher. Therefore, we adopt synthetic samples generated by BigGAN to perform optimization. We leave the study on more choices of synthetic samples in future work. 

\begin{figure*}[t]
\centering
\includegraphics[width=1.0\linewidth]{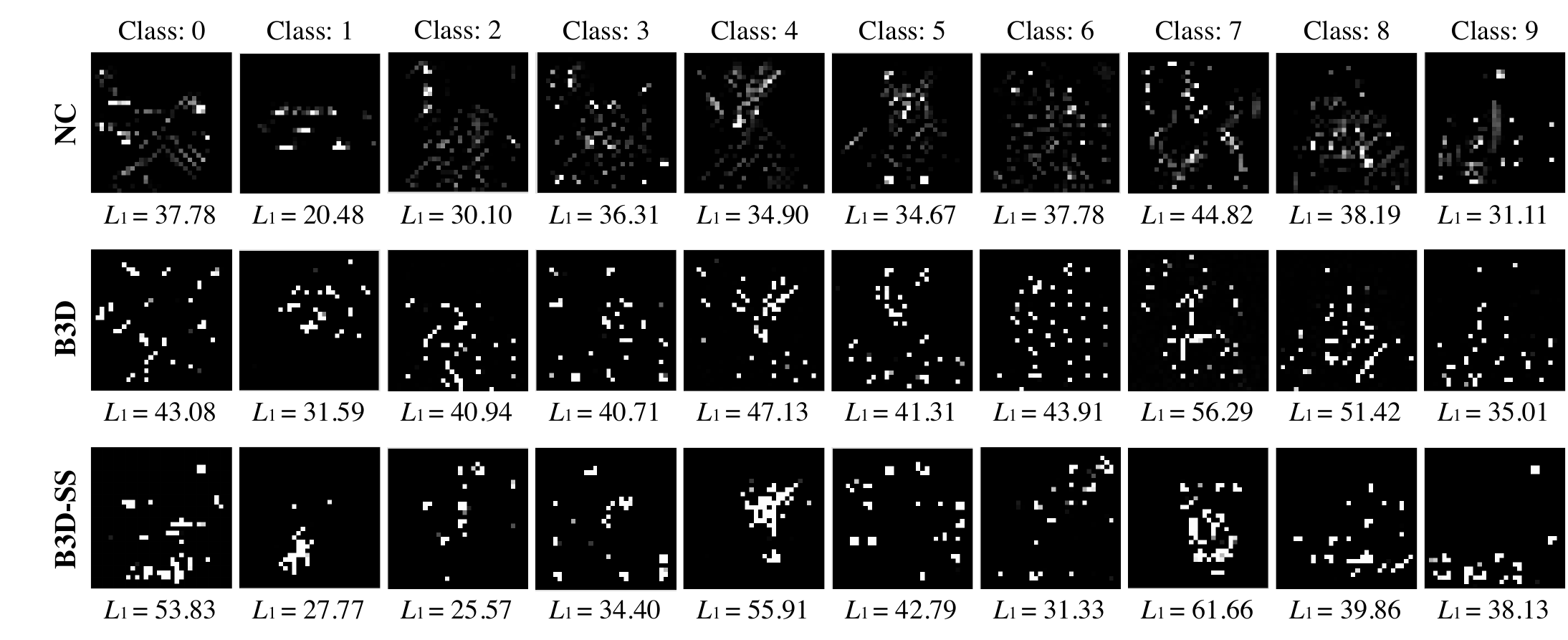}
\caption{Visualization of the reversed masks optimized by NC, B3D, and B3D-SS for all classes of a normal model on CIFAR-10. NC wrongly identifies the model as backdoored and regards class 1 to be the infected class.}
\label{fig:normal}
\end{figure*}

\section{Analysis on NC and B3D for Normal Models}\label{app:c}

In the experiments, we find that NC wrongly identifies more normal models as backdoored than B3D and B3D-SS, especially on CIFAR-10. We provide further analysis in this section. 

Fig.~\ref{fig:normal} shows an example of the wrong identification of a normal model by NC trained on CIFAR-10. Because NC relaxes the masks to be continuous in $[0,1]^d$, it can be observed that the reversed mask by NC has small amplitude but covers a large region. In this example, class 1 is identified as an infected class since the $L_1$ norm of the mask is smaller than others and is regarded as an outlier among the masks of all classes. However, this mask does not resemble the masks of true backdoor patterns. In B3D and B3D-SS, as we adopt the Bernoulli distribution to model the masks, the optimized masks tend to be close to $1$. Thus B3D and B3D-SS are less probable to optimize a mask with much smaller $L_1$ norm for a specific class. As a result, B3D and B3D-SS are less prone to this problem.

\section{Effective Positions of Backdoor Attacks}\label{app:d}

\begin{figure}[t]
\centering
\includegraphics[width=1.0\linewidth]{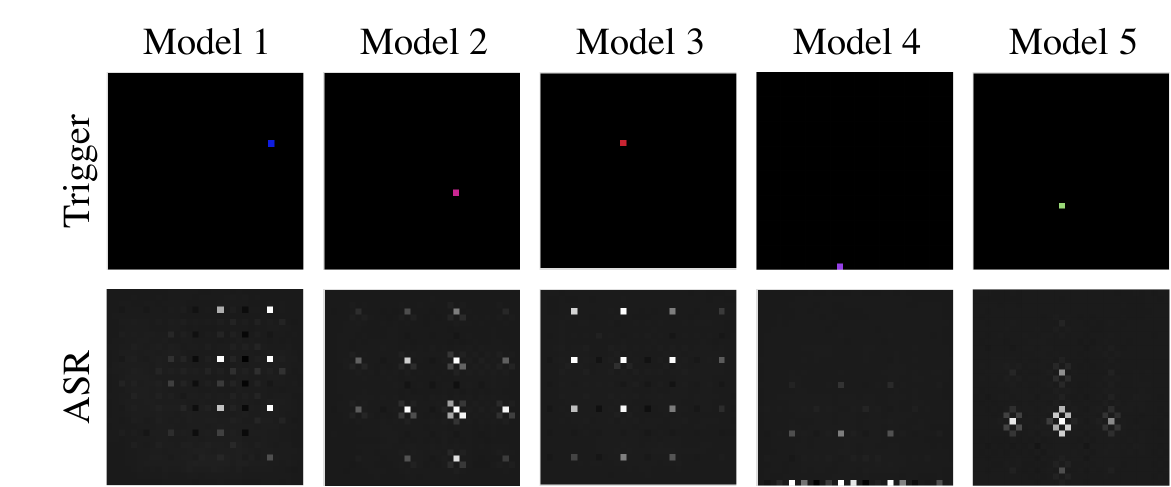}
\caption{The original triggers and the backdoor attack success rates (ASR) by applying the triggers to different positions in the input. In the second row, the value of the pixel represents the ASR at each position, \ie, a white pixel represents the $100\%$ ASR while a black pixel represents the $0\%$ ASR.}
\label{fig:position}
\end{figure}
\begin{figure}[t]
\centering
\includegraphics[width=1.0\linewidth]{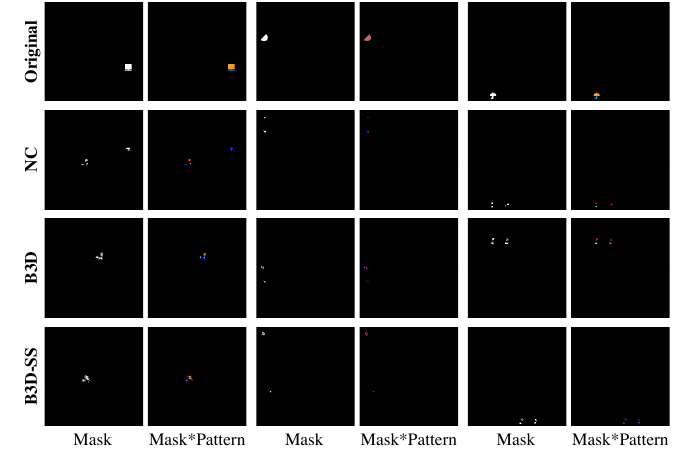}
\caption{Visualization of the original triggers and the reversed triggers optimized by NC, B3D, and B3D-SS on ImageNet.}
\label{fig:vis-imagenet}
\end{figure}

Although we typically embed a backdoor in a model at a specific input position, the reversed trigger often locates at a different position from the original trigger. We deduce that the backdoored model would learn a distribution of triggers by generalizing the original one. To validate it, we calculate the success rates of backdoor attacks by applying the trigger to all input positions.

Specifically, we randomly choose $5$ backdoored models on CIFAR-10 with $1\times1$ triggers. For each model, we insert the trigger into each position of the input and evaluate the attack success rates (ASR). We visualize the heat maps of ASR in Fig.~\ref{fig:position}. It can be seen that a lot of input positions besides the original one can induce high ASR. Thus we can conclude that the backdoored model can learn a distribution of backdoor triggers in various positions, and the backdoor detection method could converge to either one from the distribution, which does not necessarily locate at the same position as the original trigger.

\section{Visualization Results on ImageNet}\label{app:e}

We visualize the original triggers and the reversed triggers optimized by NC, B3D, and B3D-SS on ImageNet in Fig.~\ref{fig:vis-imagenet}. It can be seen that the reversed triggers do not resemble the original triggers, indicating that the backdoored models would automatically learn distinctive features from the triggers rather than remembering the exact patterns.

\begin{table*}[!t]
\small
\begin{center}
\begin{tabular}{c|c|c|c|cc|cccc}

\hline
\multirow{2}{*}{Attack} & \multirow{2}{*}{Accuracy} & \multirow{2}{*}{ASR} & \multirow{2}{*}{Method} & \multicolumn{2}{c|}{Reversed Trigger} & \multicolumn{4}{c}{Detection Results}\\
\cline{5-10}
& & & & $L_1$ norm & ASR & Case I & Case II & Case III & Case IV \\
\hline\hline
\tabincell{c}{Blended Injection} & 88.36\% & 100.00\% & \tabincell{l}{NC~\cite{wang2019neural} \\TABOR~\cite{guo2019tabor} \\ B3D (Ours) \\ B3D-SS (Ours)} & \tabincell{c}{0.499 \\ 0.640 \\ 0.865 \\ 4.320} & \tabincell{c}{98.77\% \\ 99.00\% \\ 98.99\% \\ 99.99\%} & \tabincell{c}{40/50 \\ 37/50 \\ 36/50 \\ 40/50} & \tabincell{c}{10/50 \\ 11/50 \\ 14/50 \\ 10/50} & \tabincell{c}{0/50 \\ 0/50 \\ 0/50 \\ 0/50} & \tabincell{c}{0/50 \\ 2/50 \\ 0/50 \\ 0/50} \\
\hline
\tabincell{c}{Label-Consistent}& 86.70\% & 99.92\% & \tabincell{l}{NC~\cite{wang2019neural} \\TABOR~\cite{guo2019tabor} \\ B3D (Ours) \\ B3D-SS (Ours)} & \tabincell{c}{3.092 \\ 3.291 \\ 3.737 \\ 3.783} & \tabincell{c}{98.72\% \\ 99.19\% \\ 98.92\% \\ 97.81\%} & \tabincell{c}{47/50 \\ 46/50 \\ 46/50 \\ 47/50} & \tabincell{c}{0/50 \\ 1/50 \\ 1/50 \\ 3/50} & \tabincell{c}{0/50 \\ 0/50 \\ 0/50 \\ 0/50} & \tabincell{c}{3/50 \\ 3/50 \\ 3/50 \\ 0/50} \\
\hline
\end{tabular}
\end{center}
\vspace{-3ex}
\caption{The results of backdoor detection on CIFAR-10 against the blended injection attack~\cite{chen2017targeted} and label-consistent attack~\cite{turner2019label}. We show the average accuracy and backdoor attack success rates (ASR) of the backdoored models. For the four backdoor detection methods --- NC, TABOR, B3D, and B3D-SS, we report the $L_1$ norm and attack success rates of the reversed trigger corresponding to the target class, as well as the detection results in four cases.}
\label{tab:attack-other}
\end{table*}

\begin{table*}[!t]
\small
\begin{center}
\begin{tabular}{c|c|c|c|cc|cccc}

\hline
\multirow{2}{*}{Model} & \multirow{2}{*}{Accuracy} & \multirow{2}{*}{ASR} & \multirow{2}{*}{Method} & \multicolumn{2}{c|}{Reversed Trigger} & \multicolumn{4}{c}{Detection Results}\\
\cline{5-10}
& & & & $L_1$ norm & ASR & Case I & Case II & Case III & Case IV \\
\hline\hline
Normal & 89.57\% & N/A & \tabincell{l}{NC~\cite{wang2019neural} \\TABOR~\cite{guo2019tabor} \\ B3D (Ours) \\ B3D-SS (Ours)} & \tabincell{c}{N/A \\ N/A \\ N/A \\ N/A} & \tabincell{c}{N/A \\ N/A \\ N/A \\ N/A} & \tabincell{c}{N/A \\ N/A \\ N/A \\ N/A} & \tabincell{c}{N/A \\ N/A \\ N/A \\ N/A} & \tabincell{c}{1/50 \\ 1/50 \\ 1/50 \\ 3/50} & \tabincell{c}{49/50 \\ 49/50 \\ 49/50 \\ 47/50 } \\
\hline
\tabincell{c}{Backdoored \\ ($1\times1$ trigger)} & 88.79\% & 99.64\% & \tabincell{l}{NC~\cite{wang2019neural} \\TABOR~\cite{guo2019tabor} \\ B3D (Ours) \\ B3D-SS (Ours)} & \tabincell{c}{0.980 \\ 1.014 \\ 1.085 \\ 9.247} & \tabincell{c}{98.67\% \\ 99.12\% \\ 98.81\% \\ 99.52\%} & \tabincell{c}{41/50 \\ 39/50 \\ 32/50 \\ 25/50} & \tabincell{c}{6/50 \\ 7/50 \\ 14/50 \\ 20/50} & \tabincell{c}{2/50 \\ 0/50 \\ 2/50 \\ 3/50} & \tabincell{c}{1/50 \\ 4/50 \\ 2/50 \\ 2/50} \\
\hline
\tabincell{c}{Backdoored \\ ($2\times2$ trigger)} & 88.86\% & 99.99\% & \tabincell{l}{NC~\cite{wang2019neural} \\TABOR~\cite{guo2019tabor} \\ B3D (Ours) \\ B3D-SS (Ours)} & \tabincell{c}{2.393 \\ 2.475 \\ 2.734 \\ 6.836} & \tabincell{c}{98.69\% \\ 98.98\% \\ 98.90\% \\ 99.18\%} & \tabincell{c}{46/50 \\ 43/50 \\ 41/50 \\ 31/50} & \tabincell{c}{3/50 \\ 5/50 \\ 7/50 \\ 18/50} & \tabincell{c}{1/50 \\ 0/50 \\ 2/50 \\ 1/50} & \tabincell{c}{0/50 \\ 2/50 \\ 0/50 \\ 0/50} \\
\hline
\tabincell{c}{Backdoored \\ ($3\times3$ trigger)}& 88.70\% & 100.00\% & \tabincell{l}{NC~\cite{wang2019neural} \\TABOR~\cite{guo2019tabor} \\ B3D (Ours) \\ B3D-SS (Ours)} & \tabincell{c}{3.448 \\ 3.192 \\ 3.839 \\ 5.906} & \tabincell{c}{98.60\% \\ 99.09\% \\ 98.89\% \\ 96.72\%} & \tabincell{c}{44/50 \\ 47/50 \\ 40/50 \\ 34/50} & \tabincell{c}{5/50 \\ 3/50 \\ 7/50 \\ 14/50} & \tabincell{c}{0/50 \\ 0/50 \\ 0/50 \\ 2/50} & \tabincell{c}{1/50 \\ 0/50 \\ 3/50 \\ 0/50} \\
\hline
\end{tabular}
\end{center}
\vspace{-3ex}
\caption{The results of backdoor detection on CIFAR-10 with the VGG-16 model architecture. For normal and backdoored models with different trigger sizes, we show their average accuracy and backdoor attack success rates (ASR). For the four backdoor detection methods --- NC, TABOR, B3D, and B3D-SS, we report the $L_1$ norm and attack success rates of the reversed trigger corresponding to the target class, as well as the detection results in four cases.}
\label{tab:cifar-vgg}
\end{table*}

\section{Experiments on More Settings}\label{app:f}
In this section, we provide additional experiments by considering more various backdoor attacks and training settings. The results consistently demonstrate the effectiveness of our proposed methods --- B3D and B3D-SS.
\subsection{Other Backdoor Attacks}

Besides the BadNets approach used in the main paper, we consider more backdoor attacks including the blended injection attack~\cite{chen2017targeted} and the label-consistent attack~\cite{turner2019label}.
The blended injection attack adds a $3\times3$ trigger into a random position of the image, and performs a weighted average of the original image and the trigger. The blend ratio is set as $0.2$. The poison ratio is $10\%$. We train $50$ models by the blended injection attack. The label-consistent attack does not alter the ground-truth label of the poisoned input. We adopt the adversarial manipulation approach to make the original context hard to learn, as proposed in \cite{turner2019label}. The poison ratio is $8\%$ of the whole dataset. We also train $50$ models by the label-consistent attack. 

The results of NC, TABOR, B3D, and B3D-SS against the blended injection and label-consistent attacks are shown in Table~\ref{tab:attack-other}. NC achieves $100\%$ and $94\%$ detection accuracy against the two attacks; TABOR achieves $96\%$ and $94\%$ detection accuracy; B3D achieves $100\%$ and $94\%$ detection accuracy; while B3D-SS achieves $100\%$ detection accuracy against both attacks. The results validate the effectiveness of our proposed approaches against other backdoor attacks besides BadNets.

\subsection{Different Model Architectures}

Although we study backdoor attacks and detection using the ResNet-18 model in Sec.~\ref{sec:4}, our method can generally be applied when using other model architectures. To illustrate this, we further conduct experiments on CIFAR-10 with a VGG-16~\cite{simonyan2014very} model. The experimental settings are the same as the experiments in Sec.~\ref{sec:cifar} using the ResNet-18 model, in which we also train $200$ models for evaluations.

We present the detailed results in Table~\ref{tab:cifar-vgg}. Overall, the backdoor detection accuracy achieves $98.5\%$ by NC, $96.5\%$ by TABOR, $97.0\%$ by B3D, and $97.5\%$ by B3D-SS. The results on the VGG-16 model consistently demonstrate the effectiveness of the proposed methods B3D and B3D-SS, which achieve comparable performance with NC and TABOR.

\subsection{Data Augmentation}

\begin{table}[t]
    \begin{center}\small
    \begin{tabular}{c|c|c}
    \hline
     Trigger size & Accuracy & ASR \\
     \hline\hline
     $1\times1$ & 94.68\% & 99.67\% \\
     $2\times2$ & 94.78\% & 99.99\% \\
     $3\times3$ & 95.29\% & 100.00\% \\
    \hline
    \end{tabular}
    \end{center}
    \vspace{-3ex}
    \caption{The accuracy and the backdoor attack success rates (ASR) of three backdoored models on CIFAR-10 with data augmentation.}
    \label{tab:da}
\end{table}

The previous experiments do not adopt data augmentation during training. However, data augmentation is a common technique for training DNN models. To investigate the effects of data augmentation for backdoor attacks and detection, we provide further analysis in this section.

\begin{figure}[t]
\centering
\includegraphics[width=1.0\linewidth]{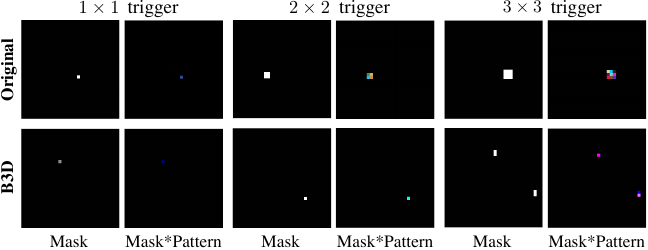}
\caption{Visualization of the original triggers and the reversed triggers optimized by B3D of three backdoored models on CIFAR-10 with data augmentation.}
\label{fig:vis-da}
\end{figure}

\begin{figure}[t]
\centering
\includegraphics[width=0.2\linewidth]{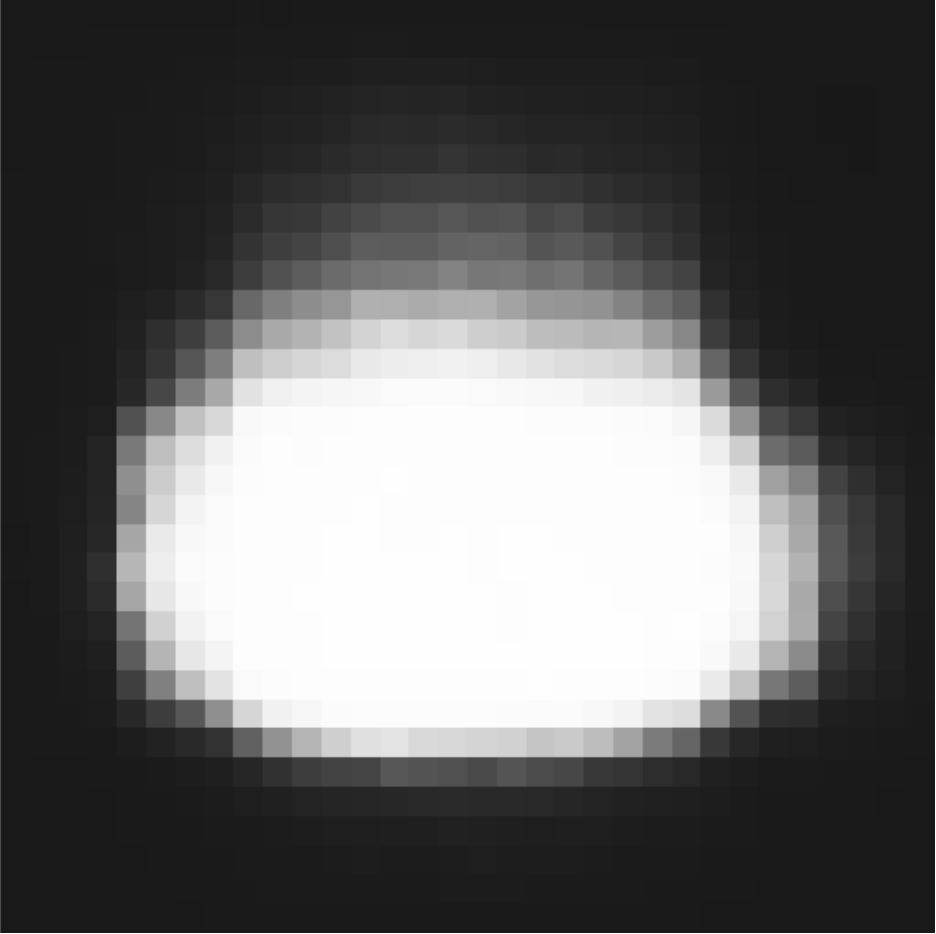}
\caption{The backdoor attack success rates (ASR) by applying the trigger to different positions in the input. We study the backdoored model using the $1\times1$ trigger on CIFAR-10 with data augmentation.}
\label{fig:position-da}
\end{figure}

We conduct experiments on CIFAR-10 with the ResNet-18 model architecture. We train one backdoored model for each trigger size of $1\times1$, $2\times2$, and $3\times3$ with data augmentation (\ie,  horizontal flips and random crops from images with $4$ pixels padded on each side). The accuracy and the backdoor attack success rates (ASR) of these models are shown in Table~\ref{tab:da}. With data augmentation, the backdoored models can achieve higher accuracy on clean test data while preserving near $100\%$ ASR for backdoor attacks. We then use B3D to perform backdoor detection of these three models. B3D successfully identifies these models as backdoored and correctly discovers the true target class. We visualize the original triggers and reversed triggers in Fig.~\ref{fig:vis-da}.

Moreover, we suspect that using data augmentation can make the effective input positions of backdoor attacks much more, because the poisoned training samples are also augmented such that the trigger will locate at many positions in the training data. Similar to the experiments in Appendix~\ref{app:c}, we use the backdoored model with the $1\times1$ trigger and show the heat map of ASR of this model in Fig.~\ref{fig:position-da}. It can be seen that the trigger is effective at a lot of positions. 

\begin{figure}[t]
\centering
\includegraphics[width=0.7\linewidth]{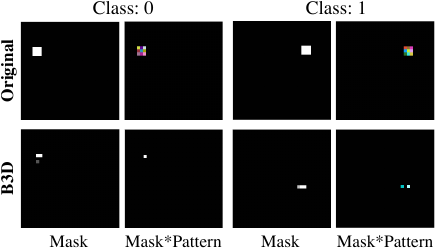}
\caption{Visualization of the original trigger and the reversed triggers optimized by B3D of a backdoored model on CIFAR-10 with two backdoors targeting at class 0 and 1.}
\label{fig:vis-two}
\end{figure}

\subsection{Multiple Infected Classes with Different Triggers}

We consider the scenario that multiple backdoors with different target classes are embedded in a model. We train a backdoored model on CIFAR-10 with two backdoors targeting at class 0 and 1, respectively. The B3D method successfully identifies both backdoors, with the reversed triggers shown in Fig.~\ref{fig:vis-two}.

\subsection{Single Infected Class with Multiple Triggers}
We consider the scenario that multiple backdoors with a single target class are embedded in a model. We train a backdoored model on CIFAR-10 with two triggers both targeting at class 0.
B3D successfully identifies the existence of backdoor attacks. 
However, we find that B3D can only restore the trigger according to one backdoor but fail to recover the trigger tied to the other. 
We think this is because that one backdoor is easier to identify than the other when we perform optimization using an objective function. It also does not harm the effectiveness of B3D in pointing out the existence of backdoored models.

\end{document}